\documentclass[prl,twocolumn,showpacs,preprintnumbers,amsmath,amssymb]{revtex4}
\usepackage{graphicx}% Include figure files
\usepackage{dcolumn}% Align table columns on decimal point
\usepackage{bm}% bold math
\usepackage{amssymb}
\usepackage{epstopdf}
\newcommand{\be}{\begin{equation}}
\newcommand{\ee}{\end{equation}}

\renewcommand{\phi}{\varphi}
\renewcommand{\epsilon}{\varepsilon}
\renewcommand{\vec}[1]{{\bf #1}}

\begin{document}
\title{
Dynamical Screening and Ferroelectric-type Excitonic Instability in Bilayer Graphene
}
\author{Rahul Nandkishore and Leonid Levitov}
\affiliation{Department of Physics, Massachusetts Institute of Technology, 77 Massachusetts Avenue, Cambridge MA02139}

\begin{abstract}
Electron interactions in undoped bilayer graphene lead to instability of the gapless state, `which-layer' symmetry breaking, and energy gap opening at the Dirac point. In contrast to single layer graphene, 
the bilayer system exhibits instability even for arbitrarily weak interaction. A controlled theory of this instability for realistic dynamically screened Coulomb interactions is developed, with full acount of dynamically generated ultraviolet cutoff. This leads to an energy gap that scales as a power law
of the interaction strength, making the excitonic instability readily observable. 
\end{abstract} 

\maketitle

Graphene, due to its unique electronic structure of a two-dimensional semimetal, provides an entirely new setting for investigating many-body phenomena \cite{r1}. Since the conduction and valence band joined together in a semimetal mimic massless Dirac particles, electronic phenomena in graphene often have direct analogs in high energy physics\cite{Gonzales94}. In particular, several authors discussed the analogy between excitonic instability in a single-layer graphene and chiral symmetry breaking in $2+1$ Quantum Electrodynamics. While this instability is absent when interactions are weak\cite{Khveschenko}, and the situation for realistic interaction strength is still debated, 
% \cite{Khveschenko,Drut,Hands}, 
the instability can be ``catalyzed'' by a magnetic field\cite{Gusynin,Herbut}. These predictions are in qualitative agreement with experiment\cite{Checkelsky}.

The effect of interactions is drastically different for semimetals with linear (type I) and quadratic (type II) electron dispersion. This was recognised in an early work \cite{Abrikosov}, where the difference in behavior was traced to the density of states at low energies, which is much lower in type I systems than in type II systems. Electronic properties of quadratically dispersing systems are governed by infrared divergences in Feynman diagrams, resulting in unconventional low energy states, whereas in linearly dispersing systems the free-particle description is robust. Indeed, excitonic instability in BLG, which has gapless quadratically dispersing electronic states, was shown to occur for arbitrarily weak short range repulsion \cite{Polini}.

In this Letter, we analyze excitonic instability in a BLG system with $1/r$ interaction, focusing on a ferroelectric (FE) state that spontaneously breaks which-layer symmetry and polarizes the layers in charge. After accounting for dynamically generated ultraviolet cutoff, treated in the RPA screening framework, we find a gap which in the weak coupling limit scales as a square of the interaction strength, $\Delta\propto (e^2/\kappa)^2$, with $e$ the electron charge and $\kappa$ the dielectric constant.
This is in contrast to the exponential BCS-like behavior of the gap expected in single layers coupled via a dielectric spacer\cite{MacDonald,Joglekar,Kharitonov}. 

\begin{figure}
\includegraphics[height = 1 in]{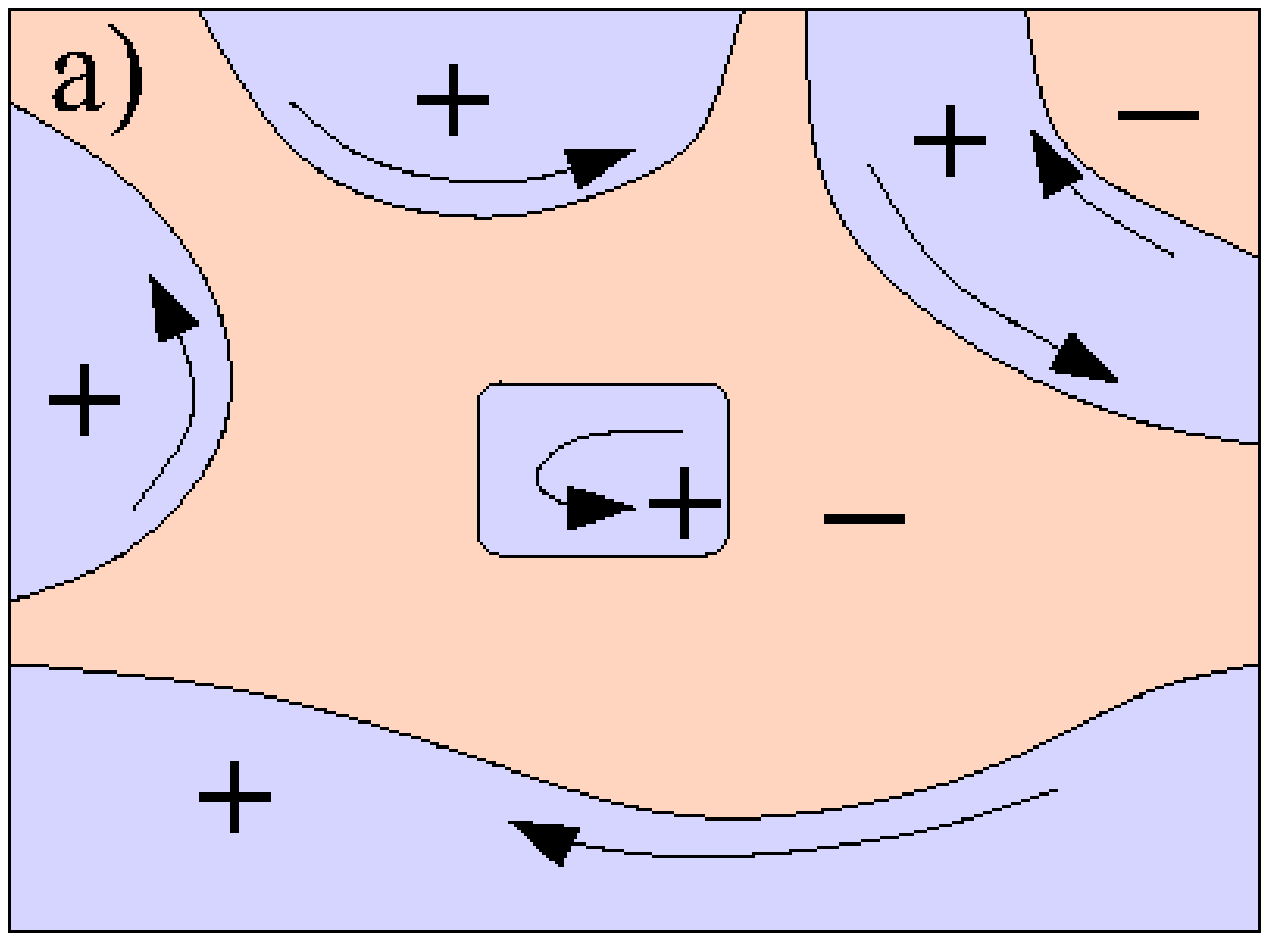}
\includegraphics[height = 1.05 in]{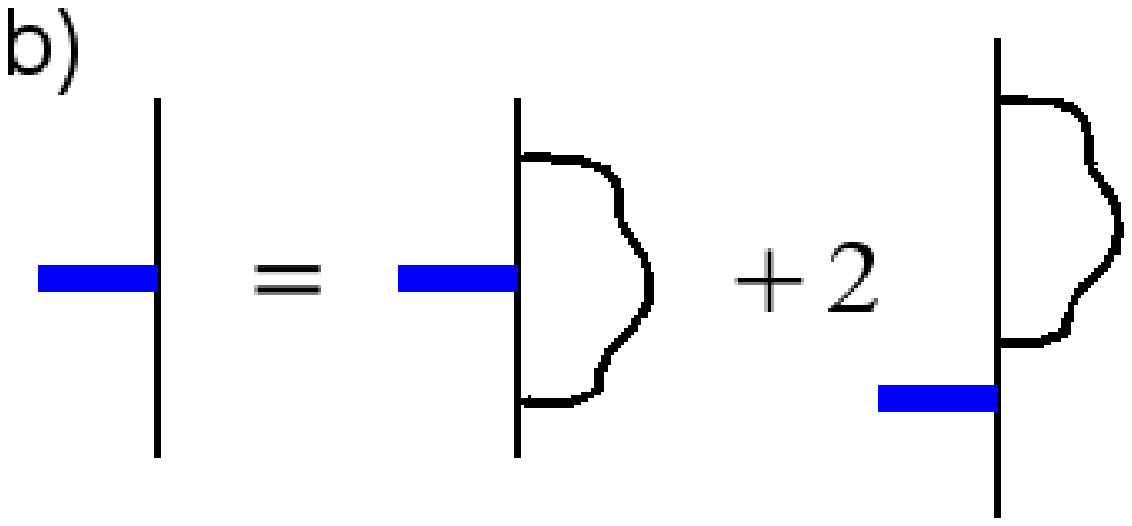}
\caption{(a) Domains of opposite polarization in the ferroelectric state. 
Valley polarized chiral edge states propagate in opposite directions along domain boundaries. (b) Diagrammatic representation of gap equation. First term is vertex correction, second term is self energy correction. Both diagrams exhibit $\log^2$ divergence which cancel to leading order. Solid, wavy and thick blue lines represent fermion propagators, the RPA interaction (\ref{eq: Ueff}), and $\Delta \tau_3$ vertex, respectively.
}
\label{fig: domains}
\vspace{-7mm}
\end{figure}

Interestingly, we find the behavior to be highly sensitive to the specifics of screening model: log divergent diagrams can become $\log^2$ divergent upon going from static to dynamic screening [see Fig.\ref{fig: domains}(b)]. Thus for a reliable estimate of the gap it is necessary to properly treat dynamic screening. We evaluate the dynamical polarization function for BLG and use it to estimate the gap value. 
% $\Delta \approx 10^{-3} me^4 /\hbar^2\kappa^2$.
For realistic parameters the predicted gap lies within the experimentally accessible range. Generalizing our approach for the anti-ferromagnetic (AF) states considered in \cite{Polini}, we find that the AF and FE states are nearly degenerate in energy.

The formation of a gapped state will manifest itself in strongly temperature dependent conductivity at  $T\lesssim\Delta$. In the presence of long-range disorder, the gapped state will occur at the p-n boundaries separating electron and hole `puddles', making these boundaries a bottleneck for transport. Hopping-like temperature dependence is indeed noted in all conductivity measurements near neutrality point in BLG but not in single-layer graphene \cite{r1}. 

Since the ferroelectric transition breaks a $Z_2$ symmetry, excitonic ordering will produce domains of opposite polarization [see Fig.\ref{fig: domains}(a)]. In the absence of disorder, the characteristic size of the domains $L$ is determined by long-range attraction between polarization in neighboring domains, in analogy with ``electronic microemulsion phases'' discussed in Ref.\cite{Spivak}, giving an estimate $8\rho^2\ln (L/a_0)\approx \Delta/a_0$, where $\rho$ is polarization density and $a_0$ is the correlation length, Eq.(\ref{eq:E_0,a_0}). 
As discussed in Ref.\cite{Morpurgo}, the boundaries between regions with opposite polarization host valley polarized edge states. Since a two-dimensional system with two domain types should exhibit percolation of edge states, the FE state should be able to carry valley currents. This could be useful for valleytronics applications. 

The low-energy Hamiltonian for BLG can be described in a `two band' approximation, neglecting the higher bands that are separated from the Dirac point by an energy gap $W \sim 0.4$ eV \cite{Falko}. The electron states are described by wavefunction taking values on the $A$ and $B$ sublattice of the upper and lower layer respectively. The non-interacting spectrum consists of quadratically dispersing quasiparticle bands $E_{\pm} = \pm p^2/2m$ with band mass $m \approx 0.054 m_e$. It is convenient to introduce the Pauli matrices $\tau_i$ that act on the sublattice space, and to define $\tau_{\pm} = \tau_1 \pm i \tau_2$ and $p_{\pm} = p_x \pm i \zeta p_y$ \cite{Falko}, where $\zeta$ = $1$ for the $K$ valley and $\zeta = -1$ for the $K'$ valley. The Hamiltonian may then be written as
\begin {eqnarray}
 \label{eq: Hamiltonian}
 H_0 &=& \sum_{\vec{p},\alpha}\psi_{\vec{p},\alpha}^\dagger\left(\frac{p_+^2}{2m} \tau_+ + \frac{p_-^2}{2m} \tau_-
\right)\psi_{\vec{p},\alpha} , \\
H &=&H_0+\frac{e^2}{2\kappa}\sum_{\rm x,x'}\frac{n(\vec x)n(\vec x')}{|\vec x-\vec x'|}
,\quad n(\vec x)=\sum_\alpha\psi_\alpha^\dagger(\vec x)\psi_\alpha(\vec x).
\nonumber
 \end{eqnarray}
The sum over $\alpha$ indicates summation over $N=4$ spin and valley species, while the dielectric constant $\kappa$ incorporates the effect of polarization of the substrate and of the higher bands of BLG.   The interaction is invariant under $SU(N)$ rotations in spin/valley space. We also approximate by treating the interlayer and intra-layer interaction as equal, and defer discussing the effect of finite layer separation until after we present our main result.

We investigate stability of the gapless state by introducing a test gap-opening perturbation $\Delta \tau_3$ into the non-interacting Hamiltonian, where $\Delta$ must be real, but may take either sign. This test perturbation explicitly breaks the $Z_2$ layer symmetry of the Hamiltonian, and corresponds to a ferroelectric instability that polarizes the layers by charge. We develop our analysis perturbatively in the interaction, and calculate the interaction renormalization of the $\Delta \tau_3$ vertex. At leading order in weak bare interactions, the vertex correction in Fig.\ref{fig: domains}(b) takes the form $\delta \Gamma = \tau_3 \delta \Delta$, where 
 \be\label{eq: vertex_correction}
\delta \Delta = - \int \frac{\Delta}{(iE+H_0)(iE - H_0)} U
.
\ee
Here $H_0$ is the Hamiltonian of the non-interacting system evaluated at $\Delta = 0$. The vertex correction is positive and preserves the form of the $\sigma_3$ vertex. Moreover, simple power counting shows that the vertex correction is divergent in the infrared for any form of interaction $U$, screened or unscreened. The infrared divergence indicates instability even for arbitrarily weak interactions (unlike monolayer graphene). The infrared divergence is power law when $U$ is the unscreened Coulomb interaction, therefore it is important to include screening even at weak coupling, to moderate the infrared divergence.

We now introduce the interaction energy scale $E_0$ and the corresponding lengthscale $a_0$, defined as  
\begin{equation}\label{eq:E_0,a_0}
E_0 = \frac{me^4}{\kappa^2\hbar^2}\approx \frac{1.47}{\kappa^2}\,{\rm eV}
, \qquad a_0 = \frac{\kappa\hbar^2}{me^2}\approx \kappa \times 1.1\,{\rm nm}
.
\end{equation}
For simplicity, we take the weak coupling limit $E_0 \lesssim W$, and neglect interaction-induced mixing of the low energy states with the higher bands [for discussion of mixing see Ref.\cite{Abergel}]. Moreover, when $U$ is the unscreened or dynamically screened Coulomb interaction, the integral in Eq.(\ref{eq: vertex_correction}) is convergent in the ultraviolet limit, without the need for any high energy cutoff. Thus $E_0$ emerges as the only energy scale in the problem. This then implies that the energy scale for the gap must scale as a power law in electric charge $\Delta \sim E_0 \sim e^4$. 

As we shall see, it is necessary to properly treat dynamic screening to obtain a reliable estimate for the gap. We therefore take $U$ in Eq.(\ref{eq: vertex_correction}) to be the dynamically screened Coulomb interaction, defined as 
\begin{equation}
\tilde U_{\omega, \vec{q}} = \frac{2\pi e^2}{\kappa q - 2\pi e^2 N \Pi_{\omega, \vec{q}}}.
\label{eq: Ueff}
\end{equation}
Here we have introduced the single species polarization function $\Pi_{\omega, \vec{q}} = - \int G(\vec{p_+}, \epsilon_+)G(\vec{p_-}, \epsilon_-) \frac{d\epsilon d^2p}{(2\pi)^3}$, where we use the notation $\vec{p}_{\pm} = \vec{p} \pm \vec{q}/2$ and $\epsilon_{\pm} = \epsilon \pm \omega/2$, and define the imaginary frequency Green function in terms of the non-interacting Hamiltonian $H_0$ as $G^{-1}(E, \vec{p}, \Delta) =  iE - H_0(\vec{p}, \Delta)$. Hence, we obtain
\begin{equation}
\label{eqn: pi delta}
\Pi_{\omega, \vec{q}, \Delta} = - 2 \int \frac{dE d^2 p}{(2\pi)^3} \frac{E_+E_- - p^2_+ p^2_- \cos(2\theta_{pq}) - \Delta^2}{(E_+^2 + p_+^4 + \Delta^2)(E_-^2 + p_-^4 + \Delta^2)}.
\end{equation}
Here, $\theta_{pq}$ is the angle between the vectors $\vec{p}_+$  and $\vec{p}_-$. To determine the dynamically screened interaction, it is sufficient to determine the polarization function in the ungapped state. We therefore suppress $\Delta$  in Eq.(\ref{eqn: pi delta}), integrate over frequencies by residues, and scale out $q$. The integral then depends on a single dimensionless parameter $\tilde \omega = 2m\omega/q^2$, and may be evaluated analytically by integrating over momenta in polar co-ordinates (see Appendix).
% \cite{supplement}.
%(details in online supplement). 
This gives an exact expression for the polarization function $\Pi_{\omega, \vec{q}, 0} = -\frac{m}{2\pi}f(\tilde \omega)$, where 
 \begin{equation}
 \label{eq: polfn}
 f(\tilde \omega)=
% \frac{2}{\xi} \arctan\xi - \frac{1}{\xi} \arctan 2\xi 
\frac{2\tan^{-1}\tilde \omega - \tan^{-1} 2\tilde \omega}{\tilde \omega}+ \ln\frac{\tilde \omega^2+1}{\tilde \omega^2+\frac14}
\approx\frac{\ln4}{\sqrt{1+u\tilde \omega^2}}
,
% \nonumber\\
%  &\approx& \frac{\ln4}{\sqrt{1+ (\frac{2\ln4}{\pi}\xi)^2}}.\nonumber
 \end{equation}
where $u=(2\ln4\pi)^2$. The right hand side provides an approximate formula that reproduces $f(\tilde \omega)$ exactly for $\tilde \omega \rightarrow 0$ and $\tilde \omega \rightarrow \infty$, interpolating accurately between the two limits. Result (\ref{eq: polfn}) agrees with the polarization function found in \cite{CastroNeto} continued to Matsubara frequencies. 

The vertex correction, Eq.(\ref{eq: vertex_correction}), calculated with the dynamically screened interaction, Eq.(\ref{eq: Ueff}), is $\log^2$ divergent in the infrared (see Appendix).
% \cite{supplement}.
%(see supplement). 
The enhanced divergence arises from the phase space region $\omega/q^2 \gg 1$, where the Coulomb interaction is not efficiently screened. However, it is also necessary to take account of the self-energy correction, so that the full gap equation is represented by Fig.\ref{fig: domains}(b). In particular, the self energy undergoes a $\log^2$ renormalization \cite{Yang}, and this can be shown to cancel the vertex correction at $\log^2$ order, leaving a residual logarithmic divergence.

\begin{figure}
\includegraphics[height = 1.3 in]{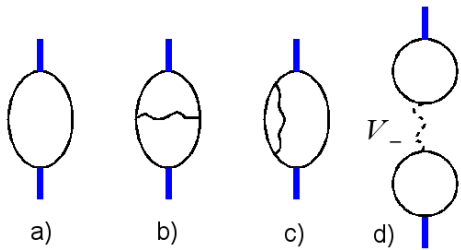}
% a)\includegraphics[height = 1.2 in]{kineticenergy.eps} \quad
% b) \includegraphics[height = 1.2 in]{exch1.eps} \quad
% c) \includegraphics[height = 1.2 in]{exch2.eps} \quad
% d) \includegraphics[height = 1.2 in]{hartree.eps}
\caption{Free energy change from gap formation at leading order in $\Delta$ and in the interaction (notation the same as in Fig.\ref{fig: domains}). The diagrams may be interpreted as: a) Kinetic energy cost from spontaneous gap opening; b, c) Interaction energy gain from gap opening; d) Hartree energy cost of layer polarization, which vanishes within the approximations of Eq.(\ref{eq: Hamiltonian}). Here $V_-$ is the difference between interlayer and intra-layer interactions. While all these diagrams are nominally $O(\Delta^2)$, $\Delta$ also appears as a logarithmic infrared cutoff in each diagram, Eq.(\ref{eqn: free energy}). }
\label{fig: free energy}
\end{figure}

Since demonstrating the cancellation of the self energy and vertex correction at leading $\log^2$ order and extracting the subleading logarithmic divergence is fairly tedious, we employ an alternative scheme for solving the gap equation Fig.\ref{fig: domains}(b). We note that calculating the free energy of BLG as a function of $\Delta$, at leading non-vanishing order in $\Delta$, simply produces the diagrams in Fig.\ref{fig: free energy}(a,b,c). Upon minimizing with respect to $\Delta$, this yields the gap equation, pictured in Fig.\ref{fig: domains}(b), with correct combinatorial coefficients. Minimizing the free energy of BLG with respect to $\Delta$ is therefore formally equivalent to solving the gap equation, and is technically simpler. It may be verified that while Fig.\ref{fig: free energy} (b) and (c) are individually $\log^3$ divergent in the infrared, their sum is only $\log^2$ divergent. This is the same leading order cancellation of divergences that is manifested by the gap equation, Fig.\ref{fig: domains}(b). 

We approximate by assuming that the gap function $\Delta$ is static and momentum independent up to energies of order $E_0$, on the grounds that the screened interaction in the particle-hole channel depends only weakly on the transferred momentum. We evaluate the kinetic energy change $\delta T$ represented by Fig.\ref{fig: free energy}(a) by including $\Delta$ in the fermion Green function. We find, with logarithmic accuracy, $\delta T = \frac{m}{2\pi} \Delta^2 \ln(\Lambda/\Delta)$, with the cutoff $\Lambda\sim E_0$.

% where $\Lambda$ is the maximum momentum scale up to which the gap persists. 

To calculate the exchange energy gain, we note that the difference in interaction energy between the gapped and ungapped states, $\delta E = E(\Delta) - E(0)$ is given, within the RPA approach, by
\begin{equation}
\label{eqn: exchange gain}
\delta E =  \int \frac{d\omega d^2 p}{(2\pi)^3} \ln\left(1 - N \tilde U_{\omega, q} (\Pi_{\omega, q, \Delta} - \Pi_{\omega, q, 0})\right).
\end{equation}
Here, $\tilde U_{\omega, q}$ is the interaction (\ref{eq: Ueff}) and $\Pi_{\omega, q, \Delta}$ is the single species polarization function in the gapped state. The problem thus reduces to that of evaluating the polarization function at finite $\Delta$.

We calculate the quantity $\Pi_\Delta - \Pi_0$ by integrating Eq.(\ref{eqn: pi delta}) over frequencies by residues, Taylor expanding to leading order in small $\Delta^2$, and analytically performing the integration over momenta, assuming as before that $\Delta$ is independent of momentum. After some algebra 
(see Appendix), we obtain
%\cite{supplement}, we obtain
% (available in the online supplement), we obtain
\begin{equation}
\Pi_\Delta - \Pi_{0} = \frac{m \Delta^2}{2\pi r^2} \left(-7 \frac{q^4}{r^2} + 4 \frac{q^8}{r^4}\right) \ln\frac{r}{\Delta}
.
\label{eqn: delta pi}
\end{equation}
We are using the notation $r^2 = \omega^2 + q^4$.

We now evaluate the exchange energy gain from gap formation by substituting Eq.(\ref{eqn: delta pi}) into Eq.(\ref{eqn: exchange gain}), and performing the integrals using polar coordinates (see Appendix).
% \cite{supplement}.
% (see supplement for details). 
Combining this with the kinetic energy cost $\delta T$, we find the free energy associated with gap opening 
\begin{equation}\label{eqn: free energy}
F(\Delta) = \frac{m}{2\pi} \Delta^2 \ln(\Lambda/\Delta) -  \frac{13 m \Delta^2}{6 \pi^3} \ln^2(N^2E_0/\Delta)
.
\end{equation}
We note the expected emergence of a natural ultraviolet cutoff. Identifying $\Lambda$ with $N^2 E_0$, and minimizing Eq.(\ref{eqn: free energy}) with respect to $\Delta$, we obtain, with logarithmic accuracy,
\begin{equation}\label{eqn: delta}
\Delta = N^2 E_0 \exp(-3\pi^2N/13) 
.
\end{equation}
We emphasize that $E_0$ appears only outside the exponential, making $\Delta$ a power law function of interaction strength at weak coupling. However, $\Delta$ is exponentially small in $N$, where $N$ is the number of fermion species participating in screening. If we had worked instead with static screening, we would have obtained $\Delta \sim N^2 E_0 \exp(-2N\ln4)$, and would have underestimated the size of the gap by an order of magnitude. 

For $N=4$ Eq.(\ref{eqn: delta}) gives $\Delta \approx 10^{-3} E_0 \approx 1.5\kappa^{-2}{\rm meV}$, upto a numerical prefactor of order unity. Meanwhile, numerically evaluating the integrals in Eq.(\ref{eqn: exchange gain}) and the kinetic energy contribution $\delta T$ [Fig.\ref{fig: free energy}(a)],
% \mpar{Do you mean $\delta T$?}
and minimizing the free energy 
%with respect to $\Delta$ 
gives $\Delta \approx 4\kappa^{-2}{\rm meV}$. This number lies within experimentally accessible range, although it can be reduced by screening in the substrate, by doping, or by disorder induced density inhomogeneity.

Thus far, we have neglected the effect of trigonal warping which leads to deviation of particle dispersion from a simple quadratic dependence, causing an overlap of the conduction and valence bands. Trigonal warping can provide an alternative low energy cutoff, preventing formation of a gapped state
if its size exceeds the estimated gap. Since the the upper estimate for trigonal warping, $1.5$ meV \cite{McCann}, is less than the 
numerical estimate $\Delta \approx 4 \kappa^{-2}{\rm meV}$, the effect of trigonal warping 
% on gap formation 
should be inessential, at least for suspended bilayers ($\kappa \approx 1$).

Our analysis can be easily generalized to any state that adds a term $\Delta \tau_3 \Omega$ to the Hamiltonian (\ref{eq: Hamiltonian}), where $\Omega$ is a $4\times 4$ hermitian matrix in spin/valley space satisfying $\Omega^2 = 1$. The FE state considered above corresponds to $\Omega=1$, wheras the AF states discussed in Ref.\cite{Polini} are characterized by $\Omega=\sigma_3\otimes 1$ or $\Omega=1\otimes \eta_3$, where $\sigma_3$ and $\eta_3$ are Pauli matrices in spin and valley space, respectively. All these inequivalent choices for $\Omega$ yield the same mean field free energy $F(\Delta)$, Eq.(\ref{eqn: free energy}), and the same gap value as was obtained for the FE state. This mean field degeneracy occurs because the Hamiltonian is invariant under $SU(4)$ spin/valley rotations, within validity of Eq.(\ref{eq: Hamiltonian}), while the states corresponding to different choices of $\Omega$ differ only in their spin/valley structure.

We now examine the effect of finite layer separation $d \approx 3 {\rm \AA}$, which differentiates the interlayer and intra-layer interactions, giving an anisotropy $V_- = \frac{1}{2}(V_{AA}-V_{AB}) = \frac{1}{2}\frac{2\pi e^2}{q} (1-e^{-qd}) = \pi e^2 d$. This anisotropy is small, because $d\ll a_0$. The leading order effect of finite layer separation is to introduce a Hartree energy for the states that polarize the layers in charge [see Fig.\ref{fig: free energy}(d)],
\begin{equation}
E_{\text{Hartree}} = \frac{N m^2}{4\pi^2} V_- \Delta^2 \ln^2(\Lambda/\Delta).
\end{equation}
It was argued in Ref.\cite{McCann} that this contribution prevents formation of the ferroelectric state. However, Ref.\cite{McCann} neglected the exchange energy. We note that the Hartree energy is of the same functional form as the exchange energy, Eq.(\ref{eqn: free energy}), but is parametrically smaller by $d/a_0 \ll 1$, and so cannot prevent the instability. 

Upon going beyond the weak coupling approximation, $d/a_0$ ceases to be a good control parameter, but our conclusions remain unchanged. This is because the $V_-$ interaction is screened as $\tilde V_-  = V_-/(1 - V_- \Pi_-)$, where $\Pi_- = \int \text{Tr} G(\epsilon_+, p_+) \tau_3 G(\epsilon_-, p_-) \tau_3 \frac{d\epsilon d^2p}{(2\pi)^3} \sim \frac{m}{2\pi} \ln(\Lambda/\Delta)$. Such logarithmic screening ensures that the Hartree energy remains smaller than the exchange energy, Eq.(\ref{eqn: free energy}), and so cannot prevent gap formation. However, when $d/a_0$ is not small, the Hartree energy may tip the balance from the FE state to one of the AF states, which do not polarize the layers by charge.

A ferromagnetic instability was predicted for unscreened interactions in \cite{CastroNeto}. However, the free energy gain from ferromagnetism was only cubic in the ferromagnetic order parameter. In contrast, the excitonic states all have a free energy gain of $O(\Delta^2)$, and should thus dominate for weak coupling.

After submitting this Letter we became aware of works \cite{MacDonald2010} and \cite{Vafek}. Instabilities in BLG are analyzed in these papers within a renormalization group framework, with interactions modeled as being short ranged. It is found that different choices of short range interaction can result in different states, gapped \cite{MacDonald2010} or gapless \cite{Vafek}.

To conclude, we have demonstrated that electron interactions drive BLG to a gapped state. For dynamically screened Coulomb interactions, taking proper account of dynamically generated ultraviolet cutoff and $\log^2$ divergences arising in this case, we obtain a gap value in experimentally accessible range ($\Delta \approx 4\kappa^{-2}{\rm meV }$).
%% an estimate for the gap of $\Delta \approx 10^{-3} E_0$, where $E_0 = m e^4/\hbar^2\kappa^2$. %, which means that even weak interactions can produce a sizeable gap. 
Manifestations of the energy gap opening in BLG at charge neutrality include temperature dependent transport at low temperatures, and also valley polarized chiral edge states propagating along domain boundaries. 

%We thank B. Feldman, J. Martin, A. Yacoby for sharing their unpublished results \cite{Yacoby}, and 
We acknowledge useful discussions with D. Abanin, P.\,A.\,Lee, Nan Gu, D. Mross, and A. Potter. This work was supported by Office of Naval Research Grant No. N00014-09-1-0724.
 %\end{acknowledgements}

% \bibliography{excitons2.bbl}

\section{Appendix A: Calculating polarization function}

Throughout this supplement, we use a shorthand whereby $p^2$ represents $p^2/2m$, unless otherwise specified. We wish to evaluate the polarisation function at non-zero $\Delta$. This was defined as
\begin{equation}
% \label{eqn: pi delta}
\Pi_{\omega, \vec{q}, \Delta} = - 2 \int \frac{dE d^2 p}{(2\pi)^3} \frac{E_+E_- - p^2_+ p^2_- \cos(2\theta_{pq}) - \Delta^2}{(E_+^2 + p_+^4 + \Delta^2)(E_-^2 + p_-^4 + \Delta^2)},
\end{equation}
where $E_{\pm} = E \pm \omega/2$, $\vec{p}_{\pm} = \vec{p} \pm \vec{q}/2$ and $\theta_{pq}$ is the angle between $\vec{p}_+$ and $\vec{p}_-$. We begin by integrating by residues over frequencies, to obtain

\begin{eqnarray}
\Pi_{\omega, q, \Delta} &=& - \int \frac{d^2p}{(2\pi)^2} \frac{\xi_+ + \xi_-}{\tilde \omega^2 + (\xi_+ + \xi_-)^2} \bigg(2 \sin(\theta_{pq})^2 - \frac{\tilde \Delta^2}{\xi_+\xi_-} \nonumber\\ &+& \cos(2\theta_{pq})(1 - \frac{\tilde p_+^2 \tilde p_-^2}{\xi_+ \xi_-})\bigg).
\end{eqnarray}
We are using the notation $\xi_{\pm}^2 = p_{\pm}^4 + \Delta^2$. Upon setting $\Delta = 0$, we obtain the polarisation function in the ungapped state. We take q to lie along the x axis without loss of generality, change to polar coordinates $p_x = p\cos{\phi}$, $p_y = p \sin \phi$ and scale out $q$. The integral then depends on a single dimensionless parameter $\tilde \omega = \omega/q^2$, and may be evaluated analytically by integrating in turn over $\phi$ and $p/q$. This gives an exact expression for the polarisation function $\Pi_{\omega, \vec{q}, 0} = -\frac{m}{2\pi}f(\tilde \omega)$, where 
 \begin{equation}
% \label{eq: polfn}
%  \Pi &=& -\frac{m}{2\pi}f(x)\\
 f(\tilde \omega)=
% \frac{2}{\xi} \arctan\xi - \frac{1}{\xi} \arctan 2\xi 
\frac{2\tan^{-1}\tilde \omega - \tan^{-1} 2\tilde \omega}{\tilde \omega}+ \ln\frac{\tilde \omega^2+1}{\tilde \omega^2+\frac14}
\approx\frac{\ln4}{\sqrt{1+u\tilde \omega^2}}
% \nonumber\\
%  &\approx& \frac{\ln4}{\sqrt{1+ (\frac{2\ln4}{\pi}\xi)^2}}.\nonumber
 \end{equation}
with $u=(2\ln4/\pi)^2$. We note that the polarisation function vanishes for $q^2 \ll \omega$, so that in this regime of parameter space, the interaction is not efficiently screened. 

We now calculate $\Pi_{\Delta} - \Pi_0$ by Taylor expanding $\Pi_{\Delta}$ to leading order in small $\Delta^2$ to obtain

\begin{widetext}
\begin{equation}
\label{eqn: e integrated}
\Pi_{\Delta} - \Pi_{0} = \Delta^2 \int \frac{d^2 p}{(2\pi)^2} \frac{p_+^2 + p_-^2}{\omega^2 + (p_+^2 + p_-^2)^2} \left( \frac{1}{p_+^2 p_-^2} - \frac12 (\frac{1}{p_+^4} + \frac{1}{p_-^4})\cos2\theta_{pq} \right) + \frac{\sin^2\theta_{pq} \left(\frac{p_+^6 + p_-^6}{p_+^2 p_-^2} + 3(p_+^2 + p_-^2) - \frac{\omega^2 (p_+^2 + p_-^2)}{p_+^2 p_-^2}\right)}{(\omega^2 + (p_+^2 + p_-^2)^2)^2}
\end{equation}
\end{widetext}

We note that the integral has singularities at $p_{\pm} = 0$, which must be regularised by introducing an IR cutoff $\Delta$. The singularities at $p_+$ and $p_-$ contribute equally, so we evaluate just the contribution from the singularities at $p_- = 0$, and multiply by two. We change variables to $\vec{p_-} \rightarrow p$, and $\vec{p_+} \rightarrow \vec{p} + \vec{q}$ and work in polar coordinates, taking $q$ to lie along the x axis. Again, we scale out q and keep only the singular terms, to obtain

\begin{equation}
\delta \Pi_{\Delta}= 2 \tilde \Delta^2 \int \frac{p dp d\varphi}{(2\pi)^2} \frac{1}{\omega^2 + 1} \frac{1}{p^2} + \frac{\sin^2\theta_{pq} \left( \frac{1}{p^2} - \frac{\omega^2}{p^2} \right)}{(\omega^2 + 1)^2} - \frac12 F(p)
\end{equation}

Here, we have measured all variables in units of $q$, have defined $\varphi$ to be the angle between $\vec{p}$ and $\vec{q}$, and have defined 

\begin{equation}
F(p) = \frac{1+2p^2 + 2p\cos\varphi}{\omega^2 + (2p^2 + 2p \cos \varphi + 1)^2}\frac{1 - 2 \sin^2\theta_{pq}}{p^4}
\end{equation}
 We can readily determine that in this basis, $\sin^2 \theta_{pq} = \sin^2\varphi / (p^2 + 2p \cos \varphi + 1)$. We can then expand $F(p)$ as a power series in small $p$. We note that the terms at $O(p^{-4})$ and $O(p^{-3})$ vanish upon angular integration, so that the leading term comes at $O(p^{-2})$. After integration over angles $\varphi$, and after making the substitution $z = p^2/2m$ we obtain
 
 \begin{eqnarray}
 \delta \Pi_{\Delta} &=& \frac{m\tilde \Delta^2}{2\pi} \left(\frac{-7}{(\omega^2 + 1)^2} + \frac{4}{(\omega^2 + 1^4)^3}\right) \int \frac{dz}{z}
 \end{eqnarray}
 After restoring $q$, we obtain
 \be
\delta \Pi_{\Delta} = \frac{m\Delta^2}{2\pi} \ln(\frac{\sqrt{\omega^2 + q^4}}{\Delta})\left( \frac{- 7 q^4}{(\omega^2 + q^4)^2} + \frac{4q^8}{(\omega^2 + q^4)^3} \right).
\label{eqn: pidelta}
 \ee
 
 \section{Appendix B: Calculating exchange energy}

In the main text, we derived 
\begin{equation}
% \label{eqn: exchange gain}
\delta E =  \int \frac{d\omega d^2 p}{(2\pi)^3} \ln\left(1 - N \tilde U_{\omega, q} (\Pi_{\omega, q, \Delta} - \Pi_{\omega, q, 0})\right).
\end{equation}

Substituting the result (\ref{eqn: pidelta}) into this, and expanding the logarithm to leading order in small $\Delta^2$, we obtain

\begin{widetext}
\begin{equation}
E_{\rm exchange} = - \int \frac{d\omega d^2q}{(2\pi)^3}  \frac{m\Delta^2}{2\pi} \ln(\frac{\sqrt{\omega^2 + q^4}}{\Delta})\left( \frac{- 7 q^4}{(\omega^2 + q^4)^2} + \frac{4q^8}{(\omega^2 + q^4)^3} \right) \frac{2\pi}{q + N q^2 \ln 4 /\sqrt{q^4 + (2\ln4/\pi)^2\omega^2}}
\end{equation}
\end{widetext}

 We have used the approximate form for the interaction, and have measured everything in units of $E_0$ and $a_0$. We note that $\Pi_{\Delta} - \Pi_0$ vanishes in the limit $q^2/\omega \ll 1$, when the Coulomb interaction is not efficiently screened. This is the region of phase space that is responsible for the vertex correction having an enhanced $\log^2$ infrared divergence (see next section). The vanishing of $\Pi_{\Delta} - \Pi_0$ in the region of phase space where the Coulomb interaction is strongest is responsible for the weakened divergence, and is another way of understanding the leading order cancellation between the vertex correction and the self energy. To formally calculate the exchange energy, we approximate the interaction by taking 
 
\begin{equation}
\frac{1}{ q + N q^2 \ln 4 /\sqrt{q^4 + (2\ln4/\pi)^2\omega^2}} \rightarrow \frac{2\theta(N - q)}{N\pi q^2/ \omega}
\end{equation}

where $\theta(x)$ represents the Heaviside step function. We take the $q^2 > \omega$ limit of the interaction, because this is where the interaction is strongest. We now substitute into Eqn.10, change to polar co-ordinates $\omega = r cos\theta$, $q^2 = r \sin \theta$, and integrate in turn over $\theta$ and $r$ to obtain the result quoted in the main text

\begin{equation}
E_{\rm exchange} = - \frac{13 m \Delta^2}{6 \pi^3} \ln^2(N^2 E_0 /\Delta)
\end{equation}
 
 \section{Appendix C: the $\log^2$ divergence of vertex correction}
 
We analyse the vertex correction $\delta \Delta$ with the dynamically screened effective interaction. After including $\Delta$ in the fermion green functions, we obtain

\begin{equation}
\delta \Delta_{\omega, q} = \int \frac{d\epsilon dz d\theta}{(2\pi)^3} \frac{\epsilon^2 + z^2 - \Delta^2_{\epsilon,p}}{(\epsilon^2 +z^2 +\Delta^2_{\epsilon,p})^2}
\frac{2\pi e^2 \Delta_{\epsilon,p}}{\kappa |\vec{p} - \vec{q}| + m e^2 N f(\tilde \omega)}
\end{equation}

We use the notation $z = p^2$ and $\tilde \omega = \omega/|\vec{p} - \vec{q}|^2$. We now
%% and $z' = |\vec{p}-\vec{q}|^2/2m$, 
measure energies and wavevectors in the units $E_0$ and $1/a_0$ respectively, to recast the vertex correction in the dimensionless form
\begin{equation}
\delta \Delta_{\omega, q} = \int \frac{d\epsilon dz d\theta}{(2\pi)^3} \frac{\epsilon^2 + z^2 - \Delta^2_{\epsilon,p}}{(\epsilon^2 +z^2 +\Delta^2_{\epsilon,p})^2}
\frac{2\pi \Delta_{\epsilon,p}}{\sqrt{2z'}+ N f(\xi)},
\label{eq: dimensionless gap equation}
\end{equation}
where $z' = (\vec{p}-\vec{q})^2$ and $\xi=(\epsilon-\omega)/z'$.
% \begin{eqnarray}
% \Delta(\omega, q) &=& \int \frac{d\epsilon dz d\theta}{(2\pi)^3} \frac{(\epsilon^2 + z^2 - \Delta^2(\epsilon,p))}{(\epsilon^2 +z^2 +\Delta^2(\epsilon,p))^2}
% \nonumber\\&\times & \frac{2\pi \Delta(\epsilon,p)}{\sqrt{2z'}+ N f((\epsilon-\omega)/z')}.
% \label{eq: dimensionless gap equation}
% \end{eqnarray}
% 
We note that integration over $z$ can be extended to infinity without running into a UV divergence.

%take $\frac{W}{E_0} \rightarrow \infty$. Now we obtain an 

We approximate by treating the gap function $\Delta$ as a constant. Using the approximate form of $f(\tilde \omega) \approx \ln 4 / \sqrt{1 + u \tilde \omega^2}$, with $u = (2\ln 4 /\pi)^2$, and changing to polar coordinates $\epsilon = \rho \sin\varphi$, $z = \rho \cos\varphi$, we integrate in turn over $\rho$ and $\varphi$. Working to leading order in small $\Delta/(N^2E_0)$, we obtain 

\begin{equation}
\delta \Delta = \Delta \frac{1}{\pi^2 N}\ln^2\left(a N^2 E_0/\Delta\right).
\end{equation}

One logarithm comes from the integration over `angles' $\varphi$, and reflects the singularity of the dynamically screened interaction in the region $q^2/\omega \ll 1$, whereas the second logarithm comes form the integration over $\rho$, and reflects the non-vanishing density of states at low energies in BLG. This second logarithm would be present even if we worked with short range (delta function) interations. 

The self energy is also $\log^2$ divergent, for the same reasons, and cancels the logarithmic divergence that comes from the singularity of the interaction as $q\rightarrow 0$. This manifests itself in the free energy calculation through $\Pi_\Delta - \Pi_0$ vanishing as $\sim q^4$ in the limit $q \rightarrow 0$. However, we note that the cancellation between vertex correction and self energy exists only to leading order. At subleading, logarithmic order, there remains an instability, as was made evident through the free energy approach. 

\end{document}